\documentclass[runningheads]{llncs}
\usepackage{graphicx}
\begin{document}
%
\title{Adversarial Prediction of Radiotherapy Treatment Machine Parameters\thanks{Supported by Elekta, Inc.}}

%
%

\author{Lyndon Hibbard}

\authorrunning{L. Hibbard}

%
%

\institute{Elekta Inc., St. Charles MO 63303, USA \\
\email{Lyn.Hibbard@elekta.com}\\
\url{http://www.elekta.com/} }

\maketitle              
\begin{abstract}
Modern external beam cancer radiotherapy applies prescribed radiation doses to tumor targets while minimally affecting nearby vulnerable organs-at-risk (OARs).  Creating a treatment plan is difficult and time-consuming with no guarantee of optimality.  Knowledge-based planning (KBP) mitigates this uncertainty by guiding planning with probabilistic models based on populations of prior clinical-quality plans.  We have developed a KBP-inspired planning model that predicts plans as realizations of the treatment machine parameters.  These are tuples of linear accelerator (Linac) gantry angles, multi-leaf collimator (MLC) apertures that shape the beam, and aperture-intensity weights that can be represented graphically in a coordinate frame isomorphic with projections (beam's-eye views) of the patient's target anatomy.  These paired data train conditional generative adversarial networks (cGANs) that estimate the MLC apertures and weights for a novel patient, thereby predicting a treatment plan.  The predicted plans' OAR sparing is close to that of the clinical plans; the predicted target coverage requires refinement to match the clinical plans' quality.  Nonetheless, the predicted plans can serve as lower bounds on plan quality, and by initializing the MLC aperture shape and weight refinement can substantially reduce the compute times for that refinement.

\end{abstract}

\section{Introduction}
Cancer radiotherapy is based on two main premises. The first is that tumor cells are less competent to repair DNA strand breaks due to radiation than nearby normal cells.  The second is that this biological difference is best exploited by irradiating the tumor with (possibly many) shaped beams of radiation performing a kind of forward tomography, matching the 3D target shape with a 3D dose.  Planning a treatment personalized for each patient requires the physician to balance tumor dose with possible damage to nearby  vulnerable organs-at-risk (OARs).  Planning difficulty is increased by anatomy complexity and by a number of dose constraints to individual targets and OARs that may be greater than the number of structures.  Maximizing plan quality by adjusting plan parameters is difficult as the effects of any change can be difficult to anticipate.  Treatment planning programs (TPPs) provide accurate models for dose physics, but they do not provide assurance that any given plan is close to the best possible, or provide direction to produce a better plan.  

To mitigate planning uncertainty, knowledge-based planning (KBP) models dose properties \cite{Kazhdan09}, \cite{Wu09}, \cite{Zhu11}, \cite{Appenzoller12} and dose distributions \cite{Shiraishi16}, \cite{McIntosh16}, \cite{Nguyen19}, \cite{Kearney20} by sampling from probabilistic models learned from populations of clinical plans.  The predicted properties must then be converted into linear accelerator (Linac)/multileaf collimator (MLC) parameters that initialize a TPP dose calculation.   

Here we report a new KBP-inspired approach that learns to predict the Linac/MLC parameters directly, bypassing intermediate steps to predict realizations of treatment plans.

\section{Methods}

\subsection{Patient Data Preparation and Treatment Planning}

This study used 178 prostate datasets, planned identically, each starting with planning CT images and anatomy contours, to obtain volume modulated arc therapy (VMAT) plans \cite{Unkelbach15}.  The cases were anonymized and the contours were curated to conform to the RTOG 0815 standard \cite{rtog0815}.  The anatomy structures included the prostate, bladder, rectum, femoral heads, anus, penile bulb, seminal vesicles and the patient external contour.  The targets are a pair of nested planning target volumes (PTVs); PTV1 enclosed the prostate plus a margin and PTV2 enclosed PTV1 and the seminal vesicles plus an additional margin.  We obtained optimal dose fluence maps using the Elekta mCycle implementation of the Erasmus iCycle program \cite{Breedveld09}, \cite{Breedveld12} following the objectives and constraints similar to the scheme in \cite{Wang16}.  The fluence maps were input to Elekta Monaco (v. 5.12) to obtain clinical, deliverable (ground truth) VMAT plans.  The goal dose for PTV2 is 60 Gy delivered over 20 fractions.  The 178 cases were divided into training (138),validation (20), and test (20) sets.

\begin{figure}
\includegraphics[width=\textwidth]{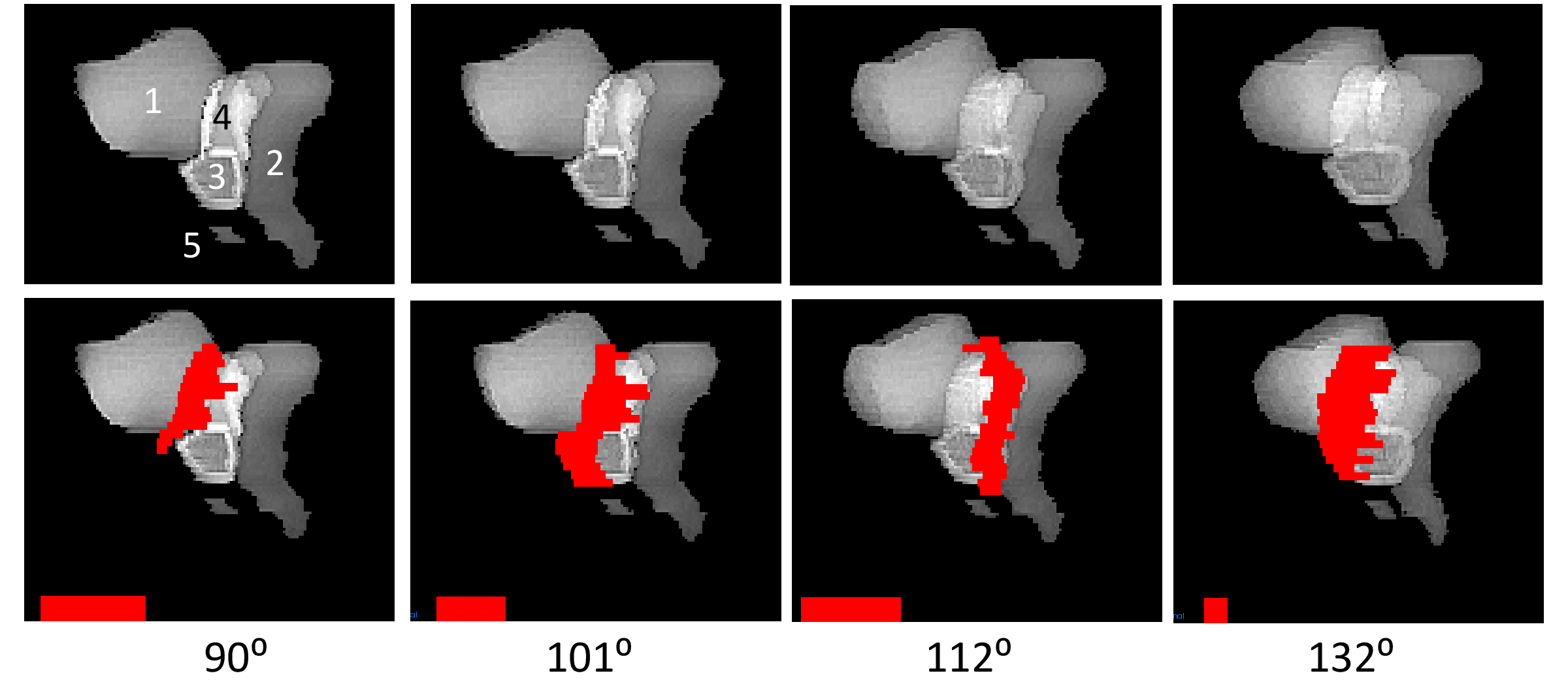}
\caption{Anatomy projections (top) and overlayed ground truth apertures (red, bottom) for a prostate target at four gantry angles.  The challenge for radiotherapy is to irradiate the target volume (prostate (3) and seminal vesicles (4) plus margins) with the prescribed dose while sparing the adjacent organs-at-risk (OARs; bladder (1), rectum (2), and penile bulb (5)).  The red bar length encodes the apertures' weights.  The combined gantry angles, apertures and aperture weights are sufficient to specify a treatment plan.  The apertures typically sweep back and forth across the target several times during one $2\pi$ arc to produce a clinical dose distribution.  See text
 for more details.} \label{fig1}
\end{figure}

The VMAT treatment is specified by the MLC apertures' shapes and their intensity weights at 100 or more discrete linac gantry angles.  The MLC creates a 2D beam profile, normal to the beam direction, by opening gaps between opposing banks of tungsten leaves.  The gantry angles, aperture leaf positions, and intensity weights are collectively the control points.  During treatment the linac gantry revolves continuously around the patient and the instantaneous linac radiation output and MLC beam shapes at any real angle are interpolated from the flanking control points.  In this way, the control points or machine parameters constitute the deliverable treatment plan.

\subsection{Data Reformatting for Supervised Learning}

The target anatomy and the VMAT control points have fundamentally different representations--anatomies are rectilinear arrays of image intensities and control points are vectors of real MLC leaf positions, plus scalar aperture weights and gantry angles.  We connect these data domains by matching graphical images of the MLC leaf apertures with beam's-eye-view projection images of the coresponding anatomy taken at the same gantry angle, and projected onto the normal plane containing the treatment isocenter (Fig. 1).  That isocenter that is the point on the gantry axis of rotation that is also the perpendicular projection of the center of the MLC.  The isocenter is also a reference point in the target volume providing a common origin for both the anatomy and treatment machine coordinate frames.    

The CT images and the structure contours were transformed into OAR and PTV 3D binary masks, weighted, summed, and reformatted to a single 3D 8-bit image in the original CT frame.  They were designed to emphasize the target volumes' edges from all directions. These 3D anatomy maps were resampled as projections using using the forward projection function of the cone beam CT reconstruction program RTK \cite{Rit13}.  Graphical images of the apertures were created from the MLC leaf positions and aligned with the plan isocenter and oriented and scaled to match the projections.  The projection and aperture images were created at 160 equi-spaced gantry angles producing one projection and one aperture image at each angle.  Each of these 2D images has dimensions 128x128 pixels (1.5 mm pixel spacing throughout) so the complete data for each patient consisted of two 128$^2$x160 volumes--one a stack of the projections and the second a stack of the aperture graphic images.  An example of this data is shown in Fig. 1.  

The graphical images also include a bar at the lower left whose length encodes the weight (cumulative meterset weight) or dose increment for that control point.  For 100 or more control points, the weights typically range in value from 0.001 to 0.01 and the space alloted for the bar length easily enables weight representation to within about 1\% of a typical weight increment.   

Finally, the ground truth plans set the collimator angle to 5$^{\circ}$ but for training the aperture graphic was depicted at 0$^{\circ}$.  After training, the test inferred apertures' leaf positions were re-written back into DICOM RT Plan files with collimator settings at 5$^{\circ}$.

%

\subsection{Deep Learning Networks and Training}

Generative adversarial networks (GANs) have become essential technologies for data modeling and inference \cite{Goodfellow14}, \cite{Goodfellow16a}.  We examine the performance of a conditional GAN (cGAN) of the pix2pix type \cite{Isola16} and a U-Net \cite{Ronneberger15}, used both as the cGAN generator network and as an independent network. Both the cGAN and the U-Net perform voxel-wise binary classification of aperture and weight-bar voxels: Is a voxel in an aperture or weight-bar or not?  (Fig. 1.)

\begin{figure}
\includegraphics[width=\textwidth]{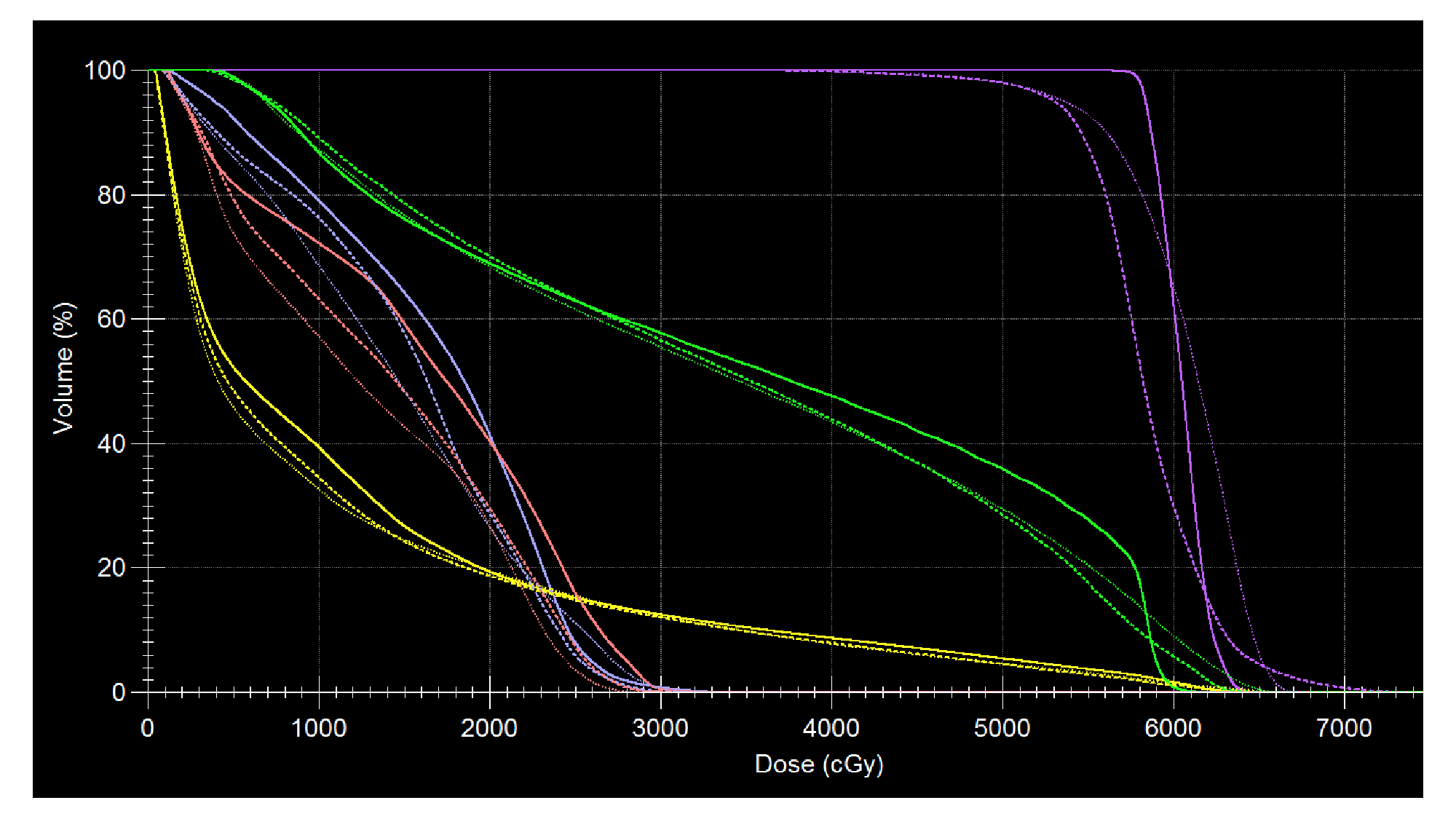}
\caption{Dose-volume histograms (DVHs) depict that fraction of a structure's voxels receiving dose greater than or equal to the dose on the horizontal axis.  Shown here are DVHs for a target and four OARs for three plans.  The plans are: clinical (solid line), 3D U-Net (heavy dashed line), and 3D cGAN (light dashed lines).  The target (PTV2) is shown at the upper right in violet.  The OARs cluster in the lower-left portion of the figure--bladder (yellow), rectum (green), left femur (blue) and right femur (orange).  See text for discussion.} \label{fig2}
\end{figure}

The networks are trained with paired, 3D data conditioned for the pix2pix cGAN as described by Isola $et al.$ \cite{Isola16}.  The cGAN discriminator network is a (70, 70, 70) PatchGAN with layer structure C16-C32-C64-C128-C128  where C is a convolution/batch normalization/LeakyReLU block with kernel (4, 4, 4), stride (2, 2, 2), Leaky ReLU slope equals 0.2, and k equals the number of filters for each Ck.  The last layer is a convolution to produce a 1D output with mean squared error loss.  

The 3D U-Net \cite{Ronneberger15}, \cite{Milletari16} generator has five levels each for the encoding and decoding branches.  The encoding branch layers are E16-E32-E48-E64-E80 where each Ek contains two convolution/ReLU/batch normalization \cite{Ioffe15} blocks with a densenet array of skip connections, kernel (3, 3, 3) and stride (1, 1, 1), and k equals the number of filters.  The decoding branch is likewise D80-D64-D48-D32-D16 where each block Dk contains a convolution transpose layer and two convolution/LeakyReLU/batch normalization blocks with kernel and stride dimensions and densenet skip connections like those of the encoding branch.  The training batch size was a single 3D volume pair.  The numbers of filters used were the largest possible given the available gpu memory.  The ADAM optimizer \cite{Kingma15} was used throughout.  The network was coded in Keras \cite{Chollet18} with TensorFlow \cite{Abadi15} and based partly on a Keras example cGAN code \cite{Brownlee19}.

\begin{figure}
\includegraphics[width=\textwidth]{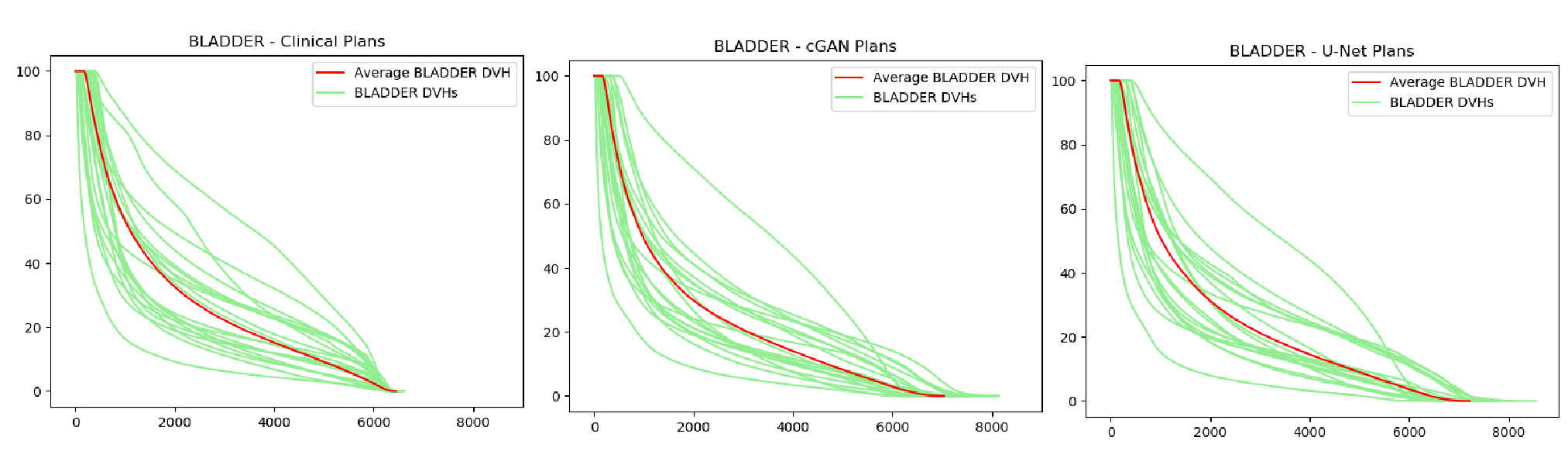}
\caption{Dose-volume histograms (DVHs) for the test population ($n=20$) for OAR bladder for three plans: ground truth clinical plans (left), cGAN predicted plans (center), and U-Net predicted plans (right).  In each panel the individual test patient's bladder DVH is depicted in green, and the group-average DVH is shown in red.} \label{fig3}
\end{figure}

\section{Results}

The training goal is a model whose inferred apertures and weights lead to an accurate dose distribution, but this calculation is impractical for all the possbile epoch-interval models.  Since MLC leaf positions (encoded by the graphic's left and right edges) dominate the weights in the TPP dose objective function, we use the Dice similarity between the deep learning model estimated apertures and the clinical plan apertures as a surrogate for plan similarity.  Plots of Dice values versus learning epoch, and visual examination of the aperture graphics helped to identify models for full studies in the TPP.  Those studies include computation of the dose distributions and DVHs for the network plans both before and after aperture and weight re-optimization, and comparisons with the corresponding clinical plans.

\begin{figure}
\includegraphics[width=\textwidth]{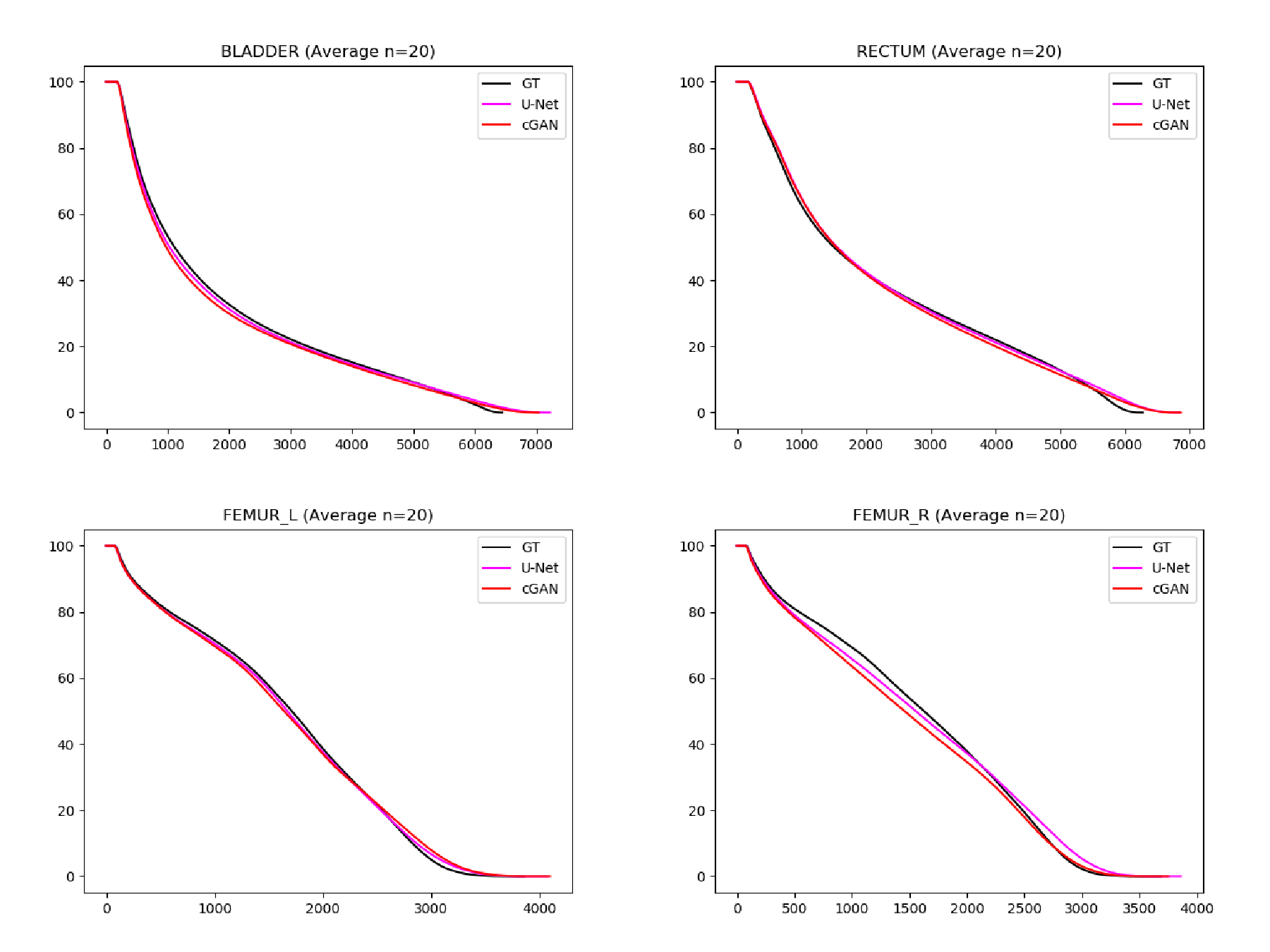}
\caption{Group-average DVHs are shown for four OARs (bladder, rectum, left femur, and right femur), in which the average DVHs for clinical plans (GT), U-Net plans, and cGAN plans are color-coded as indicated.  The average DVHs for the clinical and network-predicted plans are nearly co-incident.  See text for discussion.} \label{fig4}
\end{figure}

\subsection{Plan Quality Overview Via Dose Volume Histograms}

The cumulative dose volume histogram (DVH) depicts the fraction of a structure's voxels receiving dose greater than or equal to the dose on the horizontal axis.  Fig. 2 displays the DVHs for a target and four OARs for the three plans for a test patient. The OAR DVHs occupy the lower left of the figure and the target DVH curves are at the upper right.   The plans are the clinical (ground truth) plan (solid curves), the 3D U-Net plan (heavy dashed curves), and the 3D cGAN plan (light dashed curves).  The target is the PTV2 (violet) and the OARs are bladder (yellow), rectum (green), left femur (blue), and right femur (orange).  OAR sparing is a high priority and here the two network plans closely match the clinical plan's OAR DVHs for all four OARs, though the rectum at high dose ($>$5500 cGy) is better spared (lower volume-\% at dose) by the clinical plan.  The differences in PTV2 DVH shape indicate the predicted dose will be less homogeneous  than the clinical plan and with less-steep dose gradients.  Achieving the prescribed dose in the target with sharpest-possible dose reduction outside the target is illustrated by the clinical DVH with its sharp "shoulder" and nearly vertical descent at the prescribed dose (6000cGy).

\begin{figure}
\includegraphics[width=\textwidth]{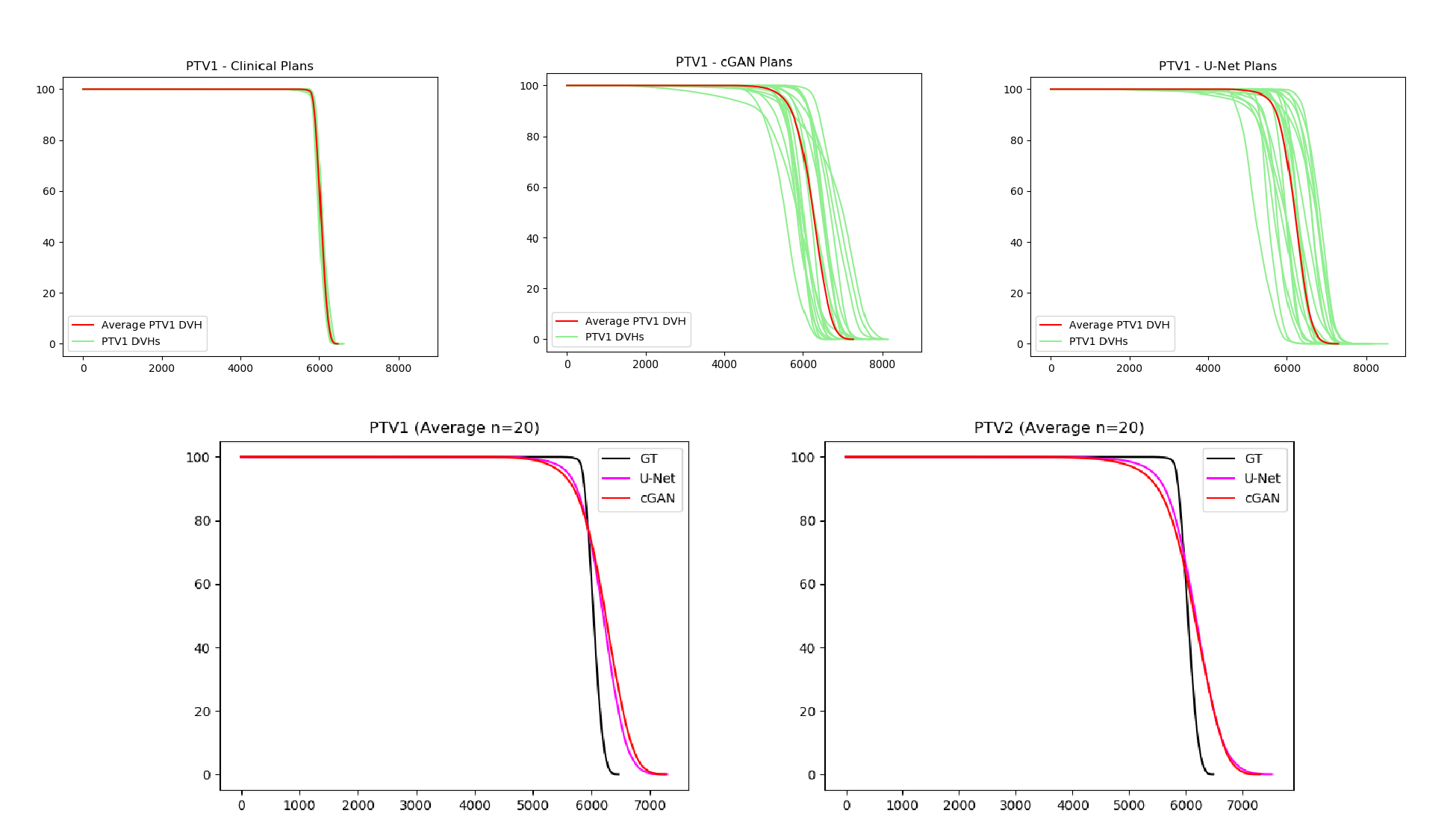}
\caption{Test patient DVHs for the target structure PTV1 are shown in the top row panels, each depicting patient DVHs (green) and the average DVH (red), for the clinical plans, the cGAN plans, and the U-Net plans, respectively.  Second row panels show the average DVHs for clinical (GT), U-Net, and cGAN plans for PTV1 (left) and PTV2 (right).  See text for discussion.} \label{fig5}
\end{figure}

\subsection{Network-Predicted OAR Sparing}

Fig. 3 shows the DVH results for a single OAR for the test population.  Three panels depict the 20 individual test patient bladder DVHs (green) and their average DVHs (red) for the clinical plans (left), the cGAN plans (middle) and the U-Net plans (right).  While the network plans and their distributions resemble that of the clinical plans, the clinical bladder DVHs show more consistent terminations at or below the 6000 cGy dose.  

Fig. 4 summarizes the group-average DVHs for the four OARs.  The four plots depict bladder (upper-left), rectum (upper-right), left femur (lower left), and right femur (lower right).  Average DVHs are color coded as indicated.  The OAR DVH averages closely match the clinical averages, except for deviations from the clinical average at high dose for some of the OARs.

\subsection{Network-Predicted Target Coverage}

Fig. 5 summarizes comparisons between the clinical and network target DVHs.  The top row plots depict the individual test patients' PTV1 DVHs (green) and the group average DVHs (red).  The target DVHs for the clinical plans (left) are strikingly more tightly clustered than those of the cGAN plans (center) or the U-Net plans (right).  The second row plots depict the average DVHs for PTV1 (left) and PTV2 (right) with color coding for the clinical (GT), U-Net, and cGAN plans indicated.  

The network predicted targets have doses similar to the clinical plans as gauged by the dose at 50\% volume, but the deviations in shape from the clinical target DVH (shape nearly a step-function) indicate that both the network plans' dose distributions are less homogeneous than the clinical plans, and that they have lower dose gradients in tissue than the clinical dose.  Thus, the network plans require a re-optimization refinement of MLC shapes and weights to be usable clinically.

\begin{figure}
\includegraphics[width=\textwidth]{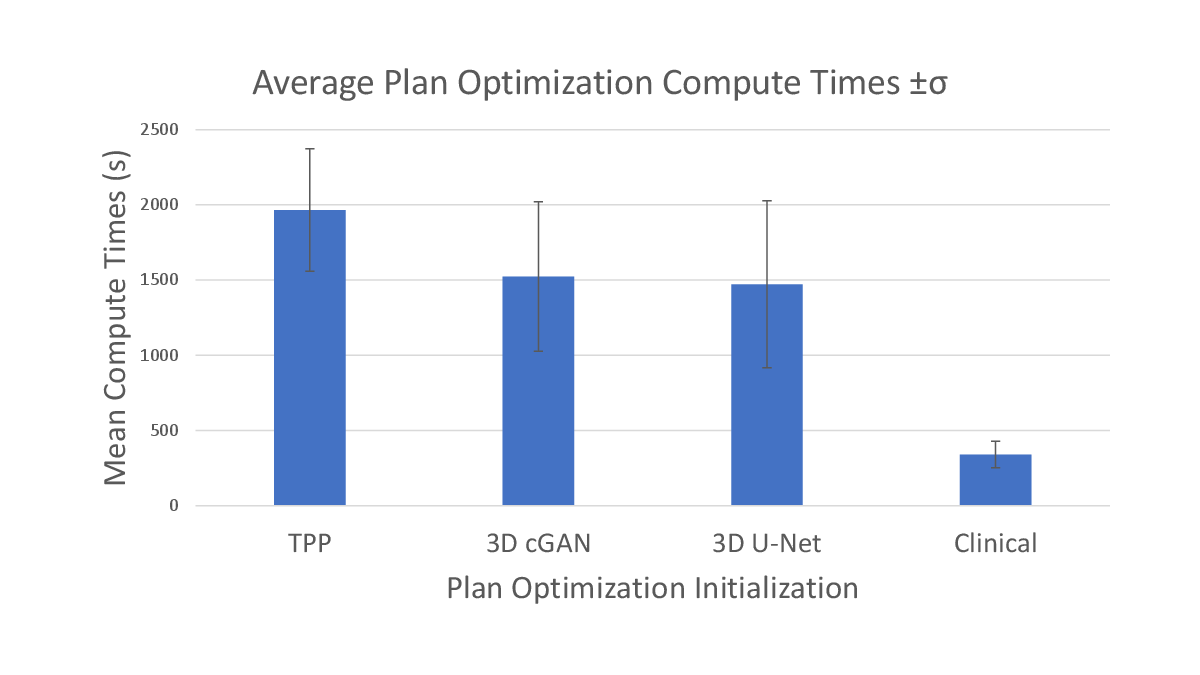}
\caption{Plan optimization adjusts the MLC apertures' shapes and weights to achieve desired target coverage and OAR sparing.  This is a bar plot of average plan optimzation compute times for the 20 test patients initialized by the TPP default method ("TPP"), the network predicted plans ("3D cGAN", "3D U-Net") and by the already-optimal clinical plans ("Clinical").  The declining trend in average times suggests that the network predicted plans are somwhat closer to the optimal plans than the TPP method, and that using the network plans to initialize plan optimization ought to take less time than default method.  See text for discussion.} \label{fig6}
\end{figure}

\subsection{Network Predicted Plans Can Accelerate Plan Optimization}

Modern treatment planning involves lengthy numerical optimizations to shape the VMAT beams to the 3D target in the patient and at the same time satisfy objectives and constraints insuring prescribed target dosage while sparing OARs.  The Elekta Monaco TPP, like other programs, creates an approximate set of VMAT control points to initiate plan optimization.  These initial control points are based on the anatomy and on assumptions about what constitutes good quality apertures--smooth, simple shapes, and a minimal number of apertures (for segments requiring multiple apertures) are typically preferred.  Plan optimization compute times depend on many factors, including (presumably) the distance in plan parameter space between the initial control point configuration and the optimum.  

Fig. 6 shows the results of timing studies involving TPP plan optimizations beginning with different starting points.  The first bar ("TPP") is the average time the Monaco TPP required for plan optimization for the 20 test patients starting with the built-in geometry-based control points (1966s).  The second and third bars ("3D cGAN" and "3D U-Net", respectively) are the average plan optimzation times for the two networks (1524s and 1472s, respectively) and the fourth bar ("Clinical") is the average time beginning with the already-optimal clinical plans (340s).  If the TPP time is taken as 100\%, the cGAN and U-Net times are 78\% and 75\%, respectively, and the clinical time is 17\%. Since the clinical time represents a lower limit to the attainable optimzation time, subtracting that time (340s) from the others yields effective reduced times of 73\% (cGAN) and 70\% (U-Net), respectively.

\section{Summary and Discussion}

This is the first report to our knowledge of a deep learning model for radiotherapy treatment plans realized as Linac/MLC control points.  The 3D cGAN and 3D U-Net networks predict these parameters for a novel patient anatomy rendered as a 3D array of beam's-eye-view projections, and these predicted apertures and their weights can be directly input into a TPP for dose calculation and plan optimization.  

Both networks' control point estimates produce OAR sparing very similar to the clinical plans, but because the network-predicted dose distributions lack sufficient  dose homogeneity and high dose gradients, control point re-optimization is needed to produce clinically useful treatment plans.    

TPPs must provide a heuristic control point initialization to begin VMAT plan computation, and we demonstrate here that the network estimated plans used to initialize plan optimization result in shorter optimzation computations.  This could have real practical benefit for TPP users.

The 3D cGAN and U-Net models can serve several functions.  The network inferences provide a lower bound to the plan quality achievable using the plan objectives and constraints for this particular radiotherapy treatment.  Thus, if a new plan in progress is not as good as the network-predicted plan, the planner knows at least that a better plan is possible, knowledge not available with current technology.  Second, the network predictions can provide a "warm start" for MLC aperture shape and weight optimzation.  Either way, deep neural network models could assist clinics without deep physics expertise or with very demanding treatment throughput requirements to achieve better efficiency and plan quality.

\section{Acknowledgements}

The author would like to thank Peter Voet, Hafid Akhiat, and Spencer Marshall for help creating the prostate plans and for many helpful discussions.

%
%
%
%

\end{document}